\documentclass[11pt]{article}
\usepackage{amssymb,amsmath,amsfonts}
\usepackage{graphicx}
\usepackage{graphics}
\usepackage{eepic,epsfig}
\usepackage{hyperref}


\begin{document}

\title{Casimir effect for fermion condensate in conical rings}
\author{A. A. Saharian$^{1}$, T. A. Petrosyan$^{1}$, A. A. Hovhannisyan$^{2}$
\\
\\
\textit{$^{1}$Department of Physics, Yerevan State University,}\\
\textit{1 Alex Manoogian Street, 0025 Yerevan, Armenia}\vspace{0.3cm}\\
\textit{$^{2}$Institute of Applied Problems of Physics NAS RA,}\\
\textit{25 Nersessian Street, 0014 Yerevan, Armenia}}
\date{}
\maketitle

\begin{abstract}
The fermion condensate (FC) is investigated for a (2+1)-dimensional massive
fermionic field confined on a truncated cone with an arbitrary planar angle
deficit and threaded by a magnetic flux. Different combinations of the
boundary conditions are imposed on the edges of the cone. They include the
bag boundary condition as a special case. By using the generalized
Abel-Plana-type summation formula for the series over the eigenvalues of the
radial quantum number, the edge-induced contributions in the FC are
explicitly extracted. The FC is an even periodic function of the magnetic
flux with the period equal to the flux quantum. Depending on the boundary
conditions, the condensate can be either positive or negative. For a
massless field the FC in the boundary-free conical geometry vanishes and the
nonzero contributions are purely edge-induced effects. This provides a
mechanism for time-reversal symmetry breaking in the absence of magnetic
fields. Combining the results for the fields corresponding to two
inequivalent irreducible representations of the Clifford algebra, the FC is
investigated in the parity and time-reversal symmetric fermionic models and
applications are discussed for graphitic cones.
\end{abstract}

\bigskip


\bigskip

\section{Introduction}

Field theoretical fermionic models in (2+1)-dimensional spacetime appear as
long-wavelength effective theories describing a relatively large class of
condensed matter systems, including graphene family materials, topological
insulators, Weyl semimetals, high-temperature superconductors, ultracold
atoms confined by lattice potentials, and nano-patterned 2D electron gases
\cite{Frad13,Mari17}. In the low-energy approximation, the corresponding
dynamics of charge carriers is governed with fairly good accuracy by the
Dirac equation, where the velocity of light is replaced by the Fermi
velocity \cite{Gusy07}-\cite{Xiao11}. The latter is much less than the
velocity of light, and this presents a unique possibility for studying
relativistic effects.

Among the most interesting topics in quantum field theory is the dependence
of the properties of the vacuum state on the geometry of the background
spacetime. The emergence of Dirac fermions in the abovementioned condensed
matter systems and availability of a number of mechanisms to control the
corresponding effective geometry provide an important opportunity to observe
different kinds of field-theoretical effects induced by the spatial geometry
and topology. In particular, it is of special interest to investigate the
influence of boundaries on the physical characteristics of the ground state.
This influence can be described by imposing appropriate boundary conditions
on the field operator. Those conditions modify the spectrum of vacuum
fluctuations and, as a consequence, the vacuum expectation values of
physical observables are shifted by an amount that depends on the bulk and
boundary geometries and on the boundary conditions. The general class of
those effects is known under the name of the Casimir effect (for reviews see
\cite{Most97}-\cite{Casi11}). In recent years, the Casimir effect for the
electromagnetic field in physical systems with graphene structures as
boundaries has been widely discussed in the literature (see references \cite%
{Bord09b}-\cite{Khus19} and references \cite{Klim09}-\cite{Klim20} for
reviews). By using external fields, different electronic phases can be
realized in Dirac materials. The magnitude and the scaling law of the
corresponding Casimir forces are essentially different for those phases \cite%
{Rodr17}. New interesting features arise in interacting fermionic systems
\cite{Eliz94b}-\cite{Flac19}.

Graphene family materials also offer a unique opportunity to investigate the
boundary-induced and topological Casimir effects for a fermionic field. On
the edges of graphene nanoribbons boundary conditions are imposed on the
effective fermionic field that ensure the zero flux of the quasiparticles.
Those conditions are sources for the Casimir-type contributions to the
expectation values of physical characteristics of the ground state.
Similarly, the periodicity conditions along compact dimensions imposed on
the fermionic field in graphene nanotubes and nanorings give rise to the
topological Casimir effect for those characteristics. As such
characteristics, in \cite{Bell09,Bell10,Bell14} the fermion condensate, the
expectation values of the current density and of the energy-momentum tensor
have been studied. The edge-induced Casimir contributions in finite length
carbon nanotubes were discussed in \cite{Bell09b}-\cite{Bell13}. Tubes with
more complicated curved geometries have been considered in \cite{Beze16}-%
\cite{Bell20b}. These geometries provide exactly solvable examples to model
the combined influence of gravity and topology on the properties of quantum
matter. Note that various mechanisms have been considered in the literature
that allow to control the effective geometry in graphene type materials \cite%
{Kole09}-\cite{Morr19}.

As background geometry, in the present paper we consider a 2-dimensional
conical space with two circular boundaries (conical ring). The corresponding
spacetime is flat and is a (2+1)-dimensional analog of the cosmic string
geometry. We investigate the influence of the edges and of the magnetic
flux, threading the ring, on the fermion condensate (FC). The corresponding
vacuum expectation values of the fermionic charge and current densities have
been recently studied in \cite{Bell20}. Among the interesting applications
of the setup under consideration are the graphitic cones. They are obtained
from a graphene sheet by cutting a sector with the angle $\pi n_{c}/3$, $%
n_{c}=1,2,\ldots ,5$, and then appropriately gluing the edges of the
remaining sector. The opening angle of the cone, obtained in this way, is
given by $\phi _{0}=2\pi (1-n_{c}/6)$. The graphitic cones with the angle $%
\phi _{0}$ for all the values corresponding to $n_{c}=1,2,\ldots ,5$, have
been observed experimentally \cite{Kris97}-\cite{Naes09}. The corresponding
electronic properties were studied in references \cite{Lamm00}-\cite{Chak11}%
. Our main interest here is the investigation of the Casimir-type
contributions to the FC induced by the edges of a conical ring for general
values of the opening angle. The ground state fermionic expectation values
for limiting cases of the geometry under consideration, corresponding to
boundary-free cones and to cones with a single circular edge, have been
examined in references \cite{Site08b}-\cite{Beze12}. In particular, the FC
has been discussed in \cite{Bell11}. The effects of finite temperature on
the FC were investigated in \cite{Bell16T,Saha19}. The formation of the FC
in models with a background scalar field has been recently discussed in \cite%
{Chu20,Chu20b}. The vacuum expectation values for the charge and current
densities on planar rings have been studied in \cite{Bell16}.

The paper is organized as follows. In the next section we describe the
geometry and present the complete set of fermionic modes. Based on those
modes, the FC is evaluated in Section \ref{sec:FC}. Various representations
are provided for the edge-induced contributions and numerical results are
presented. In Section \ref{sec:PTsym}, by combining the results for the
fields realizing two inequivalent irreducible representations of the
Clifford algebra, we consider the FC in parity and time-reversal symmetric
models. Applications to graphitic cones are discussed. The main results are
summarized in Section \ref{sec:Conc}.

\section{Geometry and the field modes}

\label{sec:Modes}

We consider a charged fermionic field in (2+1)-dimensional conical spacetime
described by the coordinates $x^{0}=t$, $x^{1}=r$, $x^{2}=\phi $, with $%
r\geqslant 0$, $0\leqslant \phi \leqslant \phi _{0}$. The corresponding
metric tensor is given by
\begin{equation}
g_{\mu \nu }=\mathrm{diag}(1,-1,-r^{2})\ .  \label{gmu}
\end{equation}%
For $\phi _{0}=2\pi $ this metric tensor corresponds to (2+1)-dimensional
Minkowski spacetime. For $\phi _{0}<2\pi $ one has a planar angle deficit $%
2\pi -\phi _{0}$ and the spacetime is flat everywhere except at the apex $%
r=0 $ where it has a delta type curvature singularity. In (2+1)-dimensional
spacetime there are two inequivalent irreducible representations of the
Clifford algebra with the $2\times 2$ Dirac matrices $\gamma _{(s)}^{\mu
}=(\gamma ^{0},\gamma ^{1},\gamma _{(s)}^{2})$, where $s=\pm 1$ correspond
to two representations. We will use the representations with $\gamma ^{0}=%
\mathrm{diag}(1,-1)$ and%
\begin{equation}
\gamma ^{1}=i\left(
\begin{array}{cc}
0 & e^{-iq\phi } \\
e^{iq\phi } & 0%
\end{array}%
\right) ,\;\gamma _{(s)}^{2}=\frac{s}{r}\left(
\begin{array}{cc}
0 & e^{-iq\phi } \\
-e^{iq\phi } & 0%
\end{array}%
\right) ,  \label{gam}
\end{equation}%
where $q=2\pi /\phi _{0}$. Note that one has the relation $\gamma
_{(s)}^{2}=-is\gamma ^{0}\gamma ^{1}/r$.

Let $\psi _{(s)}$, $s=\pm 1$, be two-component spinor fields corresponding
to two inequivalent irreducible representations of the Clifford algebra. In
the presence of an external gauge field $A_{\mu }$, the corresponding
Lagrangian density has the form
\begin{equation}
L_{(s)}=\bar{\psi}_{(s)}(i\gamma _{(s)}^{\mu }D_{(s)\mu }-m_{(s)})\psi _{(s)}
\label{Ls}
\end{equation}%
with the covariant derivative operator $D_{(s)\mu }=\partial _{\mu }+\Gamma
_{(s)\mu }+ieA_{\mu }$, the spin connection $\Gamma _{(s)\mu }$ and the
Dirac adjoint $\bar{\psi}_{(s)}=\psi _{(s)}^{\dagger }\gamma ^{0}$. We are
interested in the effects of two circular boundaries $r=a$ and $r=b$, $a<b$,
on the fermion condensate (FC)
\begin{equation}
\langle 0|\bar{\psi}_{(s)}\psi _{(s)}|0\rangle \equiv \langle \bar{\psi}%
_{(s)}\psi _{(s)}\rangle ,  \label{FCs}
\end{equation}
where $|0\rangle $ corresponds to the vacuum state. On the edges the
boundary conditions
\begin{equation}
\left( 1+i\lambda _{r}^{(s)}n_{\mu }^{(r)}\gamma _{(s)}^{\mu }\right) \psi
_{(s)}=0,\;r=a,b,  \label{BC}
\end{equation}%
will be imposed with $\lambda _{r}^{(s)}=\pm 1$ and with $n_{\mu }^{(r)}$
being the inward pointing unit vector normal to the corresponding boundary.
We can pass to the new set of fields $\psi _{(s)}^{\prime }$ defined as $%
\psi _{(+1)}^{\prime }=\psi _{(+1)}$, $\psi _{(-1)}^{\prime }=\gamma
^{0}\gamma ^{1}\psi _{(-1)}$. The corresponding Lagrangian density is
presented as $L_{(s)}=\bar{\psi}_{(s)}^{\prime }(i\gamma ^{\mu }D_{\mu
}-sm_{(s)})\psi _{(s)}^{\prime }$, where $\gamma ^{\mu }=\gamma _{(+1)}^{\mu
}$ and $D_{\mu }=D_{(+1)\mu }$. The boundary conditions are transformed to $%
\left( 1+i\lambda _{r}^{(s)\prime }n_{\mu }^{(r)}\gamma ^{\mu }\right) \psi
_{(s)}^{\prime }=0$, with $\lambda _{r}^{(s)\prime }=s\lambda _{r}^{(s)}$
and $r=a,b$. By taking into account that $\psi _{(-1)}=\gamma ^{0}\gamma
^{1}\psi _{(-1)}^{\prime }$, for the FC we get $\langle \bar{\psi}_{(s)}\psi
_{(s)}\rangle =\langle \bar{\psi}_{(s)}^{\prime }\psi _{(s)}^{\prime
}\rangle $. The boundary condition (\ref{BC}) with $\lambda _{r}^{(s)}=1$
has been used in MIT bag models to confine the quarks inside hadrons (for a
review see \cite{John75}). In condensed matter applications it is known as
infinite mass or hard wall boundary condition \cite{Berr87}. As it has been
mentioned in \cite{Berr87}, another possibility to confine the fermions
corresponds to the condition (\ref{BC}) with $\lambda _{r}^{(s)}=-1$. Note
that one has $\left( in_{\mu }^{(r)}\gamma _{(s)}^{\mu }\right) ^{2}=1$ and
for the eigenvalues of the matrix $in_{\mu }^{(r)}\gamma _{(s)}^{\mu }$ we
get $\pm 1$. Here, the upper and lower signs correspond to the boundary
conditions (\ref{BC}) with $\lambda _{r}^{(s)}=-1$ and $\lambda _{r}^{(s)}=1$%
, respectively. More general boundary conditions for the confinement of
fermions, containing additional parameters, have been discussed in \cite%
{Mcca04}-\cite{Arau19}.

\begin{figure}[tbph]
\begin{center}
\epsfig{figure=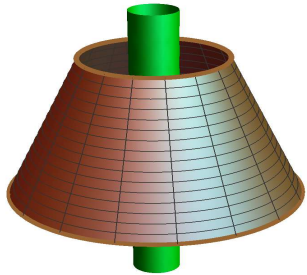,width=5cm,height=5cm}
\end{center}
\caption{The geometry of a conical ring threaded by a magnetic flux.}
\label{fig1}
\end{figure}

In the discussion below, the investigation for the FC will be presented in
terms of the fields $\psi _{(s)}^{\prime }=\psi $, omitting the prime and
the index. So, we consider a two-component fermionic field $\psi (x)$
obeying the Dirac equation
\begin{equation}
\left( i\gamma ^{\mu }D_{\mu }-sm\right) \psi (x)=0,  \label{DirEq}
\end{equation}%
and the boundary conditions%
\begin{equation}
\left( 1+i\lambda _{r}n_{\mu }^{(r)}\gamma ^{\mu }\right) \psi (x)=0,\;r=a,b,
\label{BCs}
\end{equation}%
where $\lambda _{r}=s\lambda _{r}^{(s)}$ take the values $\pm 1$. We will
consider the FC in the region $a\leqslant r\leqslant b$ where $n_{\mu
}^{(u)}=n_{u}\delta _{\mu }^{1}$, with $n_{a}=-1$ and $n_{b}=1$. Note that
the topology of the conical ring is nontrivial and the periodicity condition
on the field should be specified along the $\phi $-direction as well. Here
we impose a quasiperiodicity condition with a phase $2\pi \chi $:
\begin{equation}
\psi (t,r,\phi +\phi _{0})=e^{2\pi i\chi }\psi (t,r,\phi ).  \label{qpc}
\end{equation}%
The special cases include the untwisted and twisted fermionic fields with $%
\chi =0$ and $\chi =1/2$, respectively. For the external gauge field we
assume a simple form with the covariant components $A_{\mu }=A\delta _{\mu
}^{2}$ in the region $a\leqslant r\leqslant b$. The corresponding field
strength $F_{\mu \nu }$ vanishes on the conical ring and the effect of that
configuration of gauge field on the properties of the fermionic vacuum is
purely topological. Assuming that the 2D conical geometry under
consideration is embedded in 3D Euclidean space, the parameter $A$ can be
interpreted in terms of the magnetic flux for a gauge field $A_{k}^{\prime }$%
, $k=0,1,2,3$, living in (3+1)-dimensional flat spacetime. Introducing 3D
cylindrical coordinate system $(\rho ,\varphi ,z)$ with the axis $z$ along
the axis of the cone, we get the relations $\rho =r/q$ and $\varphi =q\phi $%
. If the magnetic field $\mathbf{B}=\mathrm{rot}\,\mathbf{A}$ corresponding
to the vector potential $A_{k}^{\prime }$ is localized in the region $\rho
<a/q$ of the 3D space, then for the magnetic flux $\Phi =\int \mathbf{B}%
\cdot d\mathbf{S}$, threading the conical ring, one obtains $\Phi =\oint
\mathbf{A}\cdot d\mathbf{l}=2\pi \rho A_{\phi }$, where as an integration
contour we have taken a circle on the conical ring and $A_{\phi }$ is the
physical azimuthal component of the vector potential. Now, by taking into
account that $A_{\phi }=-A/r$ and $2\pi \rho =\phi _{0}r$, we find $\Phi
=-\phi _{0}A$. Note that in this interpretation we have a situation similar
to that in braneworld models with extra dimensions: for a part of the fields
the whole space is accessible (the gauge field in the problem at hand) and
the another part of the fields (the fermion field) is confined to a
hypersurface (the conical ring). As an example of physical realization of
the 2D model embedded in 3D Euclidean space, in Section \ref{sec:PTsym} we
will consider graphene conical rings. The effect of the magnetic flux on the
FC, discussed below, is of the Aharonov-Bohm-type and it does not depend on
the profile of the magnetic field sourcing the flux. The spatial geometry of
the problem under consideration with the magnetic flux is presented in
Figure \ref{fig1}.

The ground state FC is expressed in terms of the fermion two-point function $%
S^{(1)}(x,x^{\prime })$ as
\begin{equation}
\left\langle \bar{\psi}\psi \right\rangle =-\lim_{x^{\prime }\rightarrow x}%
\mathrm{Tr}(S^{(1)}(x,x^{\prime })).  \label{FCdef}
\end{equation}%
The two-point function describes the correlations of the vacuum fluctuations
and is defined as the VEV\ $S_{ik}^{(1)}(x,x^{\prime })=\langle 0|[\psi
_{i}(x),\bar{\psi}_{k}(x^{\prime })]|0\rangle $ with the spinor indices $i$
and $k$. The trace in (\ref{FCdef}) is taken over those indices. The FC
plays an important role in discussions of chiral symmetry breaking and
dynamical mass generation for fermionic fields. Expanding the fermionic
operator in terms of a complete set of the positive and negative energy mode
functions $\psi _{\sigma }^{\left( +\right) }$ and $\psi _{\sigma }^{\left(
-\right) }$, obeying the conditions (\ref{BCs}), (\ref{qpc}), and using the
anticommutation relations for the fermionic annihilation and creation
operators, the following mode sum is obtained for the FC%
\begin{equation}
\left\langle \bar{\psi}\psi \right\rangle =-\frac{1}{2}\sum_{\sigma
}\sum_{\kappa =-,+}\kappa \bar{\psi}_{\sigma }^{\left( \kappa \right) }\psi
_{\sigma }^{\left( \kappa \right) }.  \label{FC}
\end{equation}%
Here, $\sigma $ stands for the complete set of quantum numbers specifying
the solutions of the equation (\ref{DirEq}), $\sum_{\sigma }$ is understood
as a summation for discrete components and as an integration for continuous
ones.

In the problem under consideration for the mode functions in the region $%
a\leqslant r\leqslant b$ one has \cite{Bell20}%
\begin{equation}
\psi _{\sigma }^{(\kappa )}=C_{\kappa }e^{iq(j+\chi )\phi -\kappa iEt}\left(
\begin{array}{c}
g_{\beta _{j},\beta _{j}}(\gamma a,\gamma r)e^{-iq\phi /2} \\
\frac{\epsilon _{j}\gamma e^{iq\phi /2}}{\kappa E+sm}g_{\beta _{j},\beta
_{j}+\epsilon _{j}}(\gamma a,\gamma r)%
\end{array}%
\right) \ ,  \label{modes}
\end{equation}%
where $E=\sqrt{\gamma ^{2}+m^{2}}$ is the energy, $j=\pm 1/2,\pm 3/2,\ldots $%
, $\epsilon _{j}=1$ for $j>-\alpha $ and $\epsilon _{j}=-1$ for $j<-\alpha $%
,
\begin{equation}
\beta _{j}=q|j+\alpha |-\epsilon _{j}/2.  \label{betj}
\end{equation}%
Here and in what follows
\begin{equation}
\alpha =\chi +eA/q=\chi -e\Phi /(2\pi ).  \label{alfa}
\end{equation}%
The radial functions in (\ref{modes}) are given by the expression%
\begin{equation}
g_{\beta _{j},\nu }(\gamma a,\gamma r)=Y_{\beta _{j}}^{(a)}(\gamma a)J_{\nu
}(\gamma r)-J_{\beta _{j}}^{(a)}(\gamma a)Y_{\nu }(\gamma r)\ ,  \label{gbet}
\end{equation}%
with the Bessel and Neumann functions $J_{\nu }(x)$, $Y_{\nu }(\gamma r)$,
and with the notation%
\begin{equation}
f_{\beta _{j}}^{(u)}\left( x\right) =\lambda _{u}n_{u}(\kappa \sqrt{%
x^{2}+m_{u}^{2}}+sm_{u})f_{\beta _{j}}\left( x\right) -\epsilon
_{j}xf_{\beta _{j}+\epsilon _{j}}\left( x\right) ,  \label{fu}
\end{equation}%
for $f=J,Y$, $u=a,b$, and $m_{u}=mu$.

The mode functions (\ref{modes}) obey the boundary condition on the edge $%
r=a $. The eigenvalues of the radial quantum number $\gamma $ are determined
by the boundary condition on the edge $r=b$. They are solutions of the
equation%
\begin{equation}
C_{\beta _{j}}(b/a,\gamma a)\equiv J_{\beta _{j}}^{(a)}\left( \gamma
a\right) Y_{\beta _{j}}^{(b)}\left( \gamma b\right) -J_{\beta
_{j}}^{(b)}\left( \gamma b\right) Y_{\beta _{j}}^{(a)}\left( \gamma a\right)
=0.  \label{Eqgam}
\end{equation}%
We will denote by $\gamma =\gamma _{l}$, $l=1,2,\ldots $, the positive roots
of this equation. The eigenvalues of $\gamma $ are expressed as $\gamma
=\gamma _{l}=z_{l}/a$. Note that under the change $(\alpha ,j)\rightarrow
(-\alpha ,-j)$ one has $\beta _{j}\rightarrow \beta _{j}+\epsilon _{j}$ and $%
\beta _{j}+\epsilon _{j}\rightarrow \beta _{j}$. From here we can see that
under the change
\begin{equation}
(\kappa ,\alpha ,j)\rightarrow (-\kappa ,-\alpha ,-j)  \label{Change}
\end{equation}%
we get
\begin{equation}
f_{\beta _{j}}^{(u)}\left( u\gamma \right) \rightarrow -\epsilon _{j}\left(
\lambda _{u}n_{u}/u\right) (\kappa E+sm)f_{\beta _{j}}^{(u)}\left( u\gamma
\right) ,  \label{Transf}
\end{equation}%
and, hence, the roots $\gamma _{l}$ are invariant under the transformation (%
\ref{Change}).

The normalization coefficient is given by
\begin{equation}
\left\vert C_{\kappa }\right\vert ^{2}=\frac{\pi qz}{16a^{2}}\frac{E+\kappa
sm}{E}T_{\beta _{j}}^{ab}(z),  \label{C+}
\end{equation}%
where $z=z_{l}=\gamma _{l}a$, $E=\sqrt{z^{2}/a^{2}+m^{2}}$, and we have
defined the function%
\begin{equation}
T_{\beta _{j}}^{ab}(z)=\frac{z}{E+\kappa sm}\left[ \frac{B_{b}J_{\beta
_{j}}^{(a)2}\left( z\right) }{J_{\beta _{j}}^{(b)2}\left( zb/a\right) }-B_{a}%
\right] ^{-1},  \label{Tab}
\end{equation}%
with
\begin{equation}
B_{u}=u^{2}\left[ E-\frac{\kappa \lambda _{u}n_{u}}{u}\left( \frac{E-\kappa
sm}{2E}+\epsilon _{j}\beta _{j}\right) \right] .  \label{Bu}
\end{equation}%
Note that the parameters $\chi $ and $\Phi $ enter in the expression of the
mode functions in the gauge-invariant combination $\alpha $. This shows that
the phase $\chi $ in the quasiperiodicity condition (\ref{qpc}) is
equivalent to a magnetic flux $-2\pi \chi /e$ threading the ring and vice
versa.

In addition to an infinite number of modes with $\gamma =\gamma _{l}$,
depending on the boundary conditions, one can have a mode with $\gamma
a=i\eta $, $\eta >0$. As it has been shown in \cite{Bell20}, for that mode $%
\eta \leqslant ma$ and, hence, $E\geqslant 0$. This means that under the
boundary conditions (\ref{BCs}) the vacuum state is always stable.

For half-integer values of the parameter $\alpha $, in addition to the modes
with $j\neq -\alpha $ and discussed above, a special mode with $j=-\alpha $
is present. The upper and lower components of the corresponding mode
functions are expressed in terms of the trigonometric functions. These mode
functions and the equation determining the eigenvalues of the radial quantum
number are given in \cite{Bell20}. For $j=-\alpha $ and for boundary
conditions with $\lambda _{b}=-\lambda _{a}$, one has also a zero energy
mode with $\gamma =im$. In a way similar to that discussed in \cite{Bell20}
for the vacuum expectation values of the charge and current densities, it
can be seen that the special mode with $j=-\alpha $ and $E\neq 0$ does not
contribute to the FC. The latter is a consequence of the cancellation of the
contributions coming from the positive and negative energy modes. The
contribution of the zero energy mode to the FC is zero as well. Note that
the latter is not the case for the expectation values of the charge and
current densities.

\section{Fermion condensate}

\label{sec:FC}

Given the complete set of fermionic modes, the FC on the conical ring is
obtained by using the mode sum formula (\ref{FC}). First let us consider the
case when all the roots of the eigenvalue equation are real. Substituting
the mode functions (\ref{modes}), the FC in the region $a\leqslant
r\leqslant b$ is presented in the form%
\begin{equation}
\left\langle \bar{\psi}\psi \right\rangle =-\frac{\pi q}{32a^{2}}%
\sum_{j}\sum_{\kappa =\pm }\sum_{l=1}^{\infty }T_{\beta _{j}}^{ab}\left(
z\right) \frac{z}{E}\left[ \left( sm+\kappa E\right) g_{\beta _{j},\beta
_{j}}^{2}\left( z,zr/a\right) +\left( sm-\kappa E\right) g_{\beta _{j},\beta
_{j}+\epsilon _{j}}^{2}\left( z,zr/a\right) \right] _{z=z_{l}},  \label{FC1}
\end{equation}%
where $E=\sqrt{z^{2}/a^{2}+m^{2}}$ and the summation goes over $j=\pm
1/2,\pm 3/2,\ldots $. The operators $\bar{\psi}$ and $\psi $ in the
left-hand side of (\ref{FC1}) are taken at the same spacetime point and the
expression on the right-hand side is divergent. Various regularization
schemes can be used to make the expression finite. To be specific, we will
assume that the regularization is done by introducing a cutoff function
without writing it explicitly. The final result for the renormalized FC does
not depend on the specific form of that function. By taking into account
that the roots $z_{l}$ are invariant under the transformation (\ref{Change})
and by using the transformation rule (\ref{Transf}) we can see that the FC
is an even periodic function of the parameter $\alpha $, defined by (\ref%
{alfa}), with the period 1. In particular, we have periodicity with respect
to the enclosed magnetic flux with the period equal to the flux quantum $%
2\pi /e$. If we present the parameter $\alpha $ in the form $\alpha
=n_{0}+\alpha _{0}$, with $|\alpha _{0}|\leqslant 1/2$ and $n_{0}$ being an
integer, then the FC will depend on the fractional part $\alpha _{0}$ only.
Note that the vacuum expectation values of the charge and current densities
are odd periodic functions of the magnetic flux with the same period.

An alternative representation of the FC is obtained from (\ref{FC1}) by
using the Abel-Plana-type formula \cite{Beze06,Saha08book}
\begin{eqnarray}
\sum_{l=1}^{\infty }w(z_{l})T_{\beta _{j}}^{ab}(z_{l}) &=&\frac{4}{\pi ^{2}}%
\int_{0}^{\infty }{dx\,}\frac{w(x)}{J_{\beta _{j}}^{(a)2}\left( x\right)
+Y_{\beta _{j}}^{(a)2}\left( x\right) }-\frac{2}{\pi }\underset{z=0}{\mathrm{%
Res}}\left[ \frac{w(z)H_{\beta _{j}}^{(1b)}\left( zb/a\right) }{C_{\beta
_{j}}(b/a,z)H_{\beta _{j}}^{(1a)}\left( z\right) }\right]  \notag \\
&&-\frac{1}{\pi }\int_{0}^{\infty }dx\,\sum_{p=+,-}\frac{w(xe^{pi\pi
/2})K_{\beta _{j}}^{(bp)}(xb/a)/K_{\beta _{j}}^{(ap)}(x)}{K_{\beta
_{j}}^{(ap)}\left( x\right) I_{\beta _{j}}^{(bp)}\left( xb/a\right)
-I_{\beta _{j}}^{(ap)}(x)K_{\beta _{j}}^{(bp)}(xb/a)}.  \label{Sum}
\end{eqnarray}%
for the function $w(z)$ analytic in the half-plane $\mathrm{Re}\,z>0$ of the
complex plane $z$. Here and below $H_{\nu }^{(l)}(x)$, with $l=1,2$, are the
Hankel functions and the notation $H_{\beta _{j}}^{(lu)}\left( x\right) $ is
defined in accordance with (\ref{fu}). In the second integral on the
right-hand side of (\ref{Sum}), for the modified Bessel functions $f_{\nu
}(x)=I_{\nu }(x),K_{\nu }(x)$, the notations
\begin{equation}
f_{\beta _{j}}^{(up)}(x)=\delta _{f}xf_{\beta _{j}+\epsilon _{j}}\left(
x\right) +\lambda _{u}n_{u}\left[ \kappa \sqrt{\left( xe^{p\pi i/2}\right)
^{2}+m_{u}^{2}}+sm_{u}\right] f_{\beta _{j}}\left( x\right) ,  \label{fjp}
\end{equation}%
are introduced with $p=+,-$, $u=a,b$, and
\begin{equation}
\delta _{I}=1,\;\delta _{K}=-1.  \label{delIK}
\end{equation}%
Additional conditions on the function $w(z)$ are given in \cite{Saha08book}.
By taking into account that
\begin{equation}
\sqrt{\left( xe^{p\pi i/2}\right) ^{2}+m_{u}^{2}}=\left\{
\begin{array}{cc}
\sqrt{m_{u}^{2}-x^{2}}, & x<m_{u}, \\
pi\sqrt{x^{2}-m_{u}^{2}}, & x>m_{u},%
\end{array}%
\right.  \label{sqroot}
\end{equation}%
for $x\geqslant 0$, we see that $f_{\beta _{j}}^{(u+)}(x)=f_{\beta
_{j}}^{(u-)}(x)$ in the range $x\in $ $[0,m_{u}]$. In addition, for the
function $w(x)$ corresponding to the series over $l$ in (\ref{FC1}) one has $%
w(xe^{-i\pi /2})=-w(xe^{i\pi /2})$ for $x\in $ $[0,m_{a}]$. From these
properties it follows that for the FC the integrand of the last integral in (%
\ref{Sum}) vanishes in the integration range $x\in $ $[0,m_{a}]$. It can
also be seen that for the FC the residue in (\ref{Sum}) is zero.

As a result, applying the formula (\ref{Sum}) for the series over $l$ in (%
\ref{FC1}), the FC in the region $a\leqslant r\leqslant b$ is decomposed
into two contributions. The first one, denoted below as $\left\langle \bar{%
\psi}\psi \right\rangle _{a}$, comes from the first term in the right-hand
side of (\ref{Sum}) and is presented in the form%
\begin{equation}
\left\langle \bar{\psi}\psi \right\rangle _{a}=-\frac{q}{8\pi a^{2}}%
\sum_{j}\sum_{\kappa =\pm }\int_{0}^{\infty }dz\,\frac{z}{E}\frac{\left(
sm+\kappa E\right) g_{\beta _{j},\beta _{j}}^{2}\left( z,zr/a\right) +\left(
sm-\kappa E\right) g_{\beta _{j},\beta _{j}+\epsilon _{j}}^{2}\left(
z,zr/a\right) }{J_{\beta _{j}}^{(a)2}(z)+Y_{\beta _{j}}^{(a)2}(z)}.
\label{FCa1}
\end{equation}%
The second contribution comes from the last term in (\ref{Sum}). Introducing
the modified Bessel functions, we get the following representation of the
FC:
\begin{eqnarray}
\left\langle \bar{\psi}\psi \right\rangle &=&\left\langle \bar{\psi}\psi
\right\rangle _{a}+\frac{q}{2\pi ^{2}}\sum_{n=0}^{\infty }\sum_{p=\pm
1}\int_{m}^{\infty }dx\frac{x}{\sqrt{x^{2}-m^{2}}}\mathrm{Re}\left\{ \frac{%
K_{n_{p}}^{(b)}\left( bx\right) /K_{n_{p}}^{(a)}\left( ax\right) }{%
G_{n_{p}}^{(ab)}(ax,bx)}\right.  \notag \\
&&\times \left. \left[ (sm+i\sqrt{x^{2}-m^{2}})G_{n_{p},n_{p}}^{\left(
a\right) 2}\left( ax,rx\right) -(sm-i\sqrt{x^{2}-m^{2}})G_{n_{p},n_{p}+1}^{%
\left( a\right) 2}\left( ax,rx\right) \right] \right\} ,  \label{FC2}
\end{eqnarray}%
where instead of the summation over $j$ we have introduced the summation
over $n$ with%
\begin{equation}
n_{p}=q(n+1/2+p\alpha _{0})-1/2.  \label{np}
\end{equation}%
Here and in what follows, for the functions $f_{\nu }\left( z\right) =I_{\nu
}\left( z\right) ,K_{\nu }\left( z\right) $ we use the notation%
\begin{equation}
f_{n_{p}}^{(u)}(z)=\delta _{f}zf_{n_{p}+1}\left( z\right) +\lambda
_{u}n_{u}(i\sqrt{z^{2}-m_{u}^{2}}+sm_{u})f_{n_{p}}\left( z\right) ,
\label{fu2}
\end{equation}%
with $u=a,b$, and the functions in the right-hand side of (\ref{FC2}) are
defined by
\begin{eqnarray}
G_{n_{p},\nu }^{(u)}(x,y) &=&K_{n_{p}}^{(u)}\left( x\right) I_{\nu }\left(
y\right) -(-1)^{\nu -n_{p}}I_{n_{p}}^{(u)}\left( x\right) K_{\nu }\left(
y\right) ,  \notag \\
G_{n_{p}}^{(ab)}(x,y) &=&K_{n_{p}}^{(a)}\left( x\right)
I_{n_{p}}^{(b)}\left( y\right) -I_{n_{p}}^{(a)}\left( x\right)
K_{n_{p}}^{(b)}\left( y\right) .  \label{Gen}
\end{eqnarray}%
Similar to the case of the charge and current densities, discussed in \cite%
{Bell20}, it can be shown that the expression (\ref{FC2}) is valid in the
presence of bound states as well. Under the replacements $\lambda
_{u}\rightarrow -\lambda _{u}$, $s\rightarrow -s$ we have $%
f_{n_{p}}^{(u)}(z)\rightarrow \left[ f_{n_{p}}^{(u)}(z)\right] ^{\ast }$ and
the last term in (\ref{FC2}) changes the sign.

The term $\left\langle \bar{\psi}\psi \right\rangle _{a}$ in (\ref{FC2})
does not depend on $b$. By using the asymptotic formulas for the modified
Bessel functions (see, for example, \cite{Abra72}), it can be seen that in
the limit $b\rightarrow \infty $ the last term in (\ref{FC2}) behaves as $%
e^{-2mb}$ for a massive field and like $\left( a/b\right) ^{q(1-2|\alpha
_{0}|)+1}$ for a massless field. From here it follows that $\left\langle
\bar{\psi}\psi \right\rangle _{a}=\lim_{b\rightarrow \infty }\left\langle
\bar{\psi}\psi \right\rangle $ and the contribution $\left\langle \bar{\psi}%
\psi \right\rangle _{a}$ presents the FC in the region $a\leqslant r<\infty $
of (2+1)-dimensional conical spacetime for a fermionic field obeying the
boundary condition (\ref{BCs}) at $r=a$. Hence, the last term in (\ref{FC2})
is interpreted as the contribution induced by the second boundary at $r=b$
when we add it to the conical geometry with a single edge at $r=a$. In order
to further extract the edge-induced contribution in $\left\langle \bar{\psi}%
\psi \right\rangle _{a}$ we use the relation%
\begin{equation}
\frac{g_{\beta _{j},\nu }^{2}(z,y)}{J_{\beta _{j}}^{(a)2}(z)+Y_{\beta
_{j}}^{(a)2}(z)}=J_{\nu }^{2}(y)-\sum_{l=1,2}\frac{J_{\beta
_{j}}^{(a)}(z)H_{\nu }^{(l)2}(y)}{2H_{\beta _{j}}^{(la)}(z)}.  \label{Id1}
\end{equation}%
where $\nu =\beta _{j},\beta _{j}+\epsilon _{j}$. This relation is easily
obtained by taking into account that $J_{\beta _{j}}^{(a)2}(z)+Y_{\beta
_{j}}^{(a)2}(z)=H_{\beta _{j}}^{(1a)}(z)H_{\beta _{j}}^{(2a)}(z)$. Applying (%
\ref{Id1}) for separate terms in (\ref{FCa1}), we can see that the part in
the FC coming from the first term in the right-hand side of (\ref{Id1}),
denoted here by $\left\langle \bar{\psi}\psi \right\rangle _{0}$, does not
depend on $a$ and is presented as%
\begin{equation}
\left\langle \bar{\psi}\psi \right\rangle _{0}=-\frac{qsm}{4\pi }%
\sum_{j}\int_{0}^{\infty }dx\,x\frac{J_{\beta _{j}}^{2}\left( xr\right)
+J_{\beta _{j}+\epsilon _{j}}^{2}\left( xr\right) }{\sqrt{x^{2}+m^{2}}}.
\label{FC0}
\end{equation}%
This part corresponds to the FC in a boundary-free conical space and has
been investigated in \cite{Bell11} for the case $s=1$. The corresponding
renormalized value is given by the expression
\begin{eqnarray}
&&\langle \bar{\psi}\psi \rangle _{0,\mathrm{ren}}=-\frac{sm}{2\pi r}\Big\{%
\sum_{l=1}^{[q/2]}(-1)^{l}\frac{\cot (\pi l/q)}{e^{2mr\sin (\pi l/q)}}\cos
(2\pi l\alpha _{0})  \notag \\
&&\qquad +\frac{q}{\pi }\sum_{\delta =\pm 1}\cos \left[ q\pi \left(
1/2+\delta \alpha _{0}\right) \right] \int_{0}^{\infty }dy\frac{\tanh y}{%
e^{2mr\cosh y}}\frac{\sinh [q\left( 1-2\delta \alpha _{0}\right) y]}{\cosh
(2qy)-\cos (q\pi )}\Big\},  \label{FC0ren2}
\end{eqnarray}%
where $[q/2]$ is the integer part of $q/2$. Note that for points away from
the edges of the conical ring the boundary-induced contribution in the FC is
finite and the renormalization is required for the boundary-free part only.

The contribution to the FC $\left\langle \bar{\psi}\psi \right\rangle _{a}$
coming from the last term in (\ref{Id1}) is induced by the edge at $r=a$ in
the region $a\leqslant r<\infty $. That contribution is further transformed
by rotating the integration contour over $z$ by the angle $\pi /2$ for the
terms with the Hankel functions $H_{\beta _{j}}^{(1)}(zr/a)$, $H_{\beta
_{j}+\epsilon _{j}}^{(1)}(zr/a)$, and by the angle $-\pi /2$ for the terms
with the functions $H_{\beta _{j}}^{(2)}(zr/a)$, $H_{\beta _{j}+\epsilon
_{j}}^{(2)}(zr/a)$. The parts of the integrals over the intervals $[0,ima]$
and $[0,-ima]$ cancel each other. Introducing in the remaining integrals the
modified Bessel functions the FC $\left\langle \bar{\psi}\psi \right\rangle
_{a}$ is presented in the form%
\begin{eqnarray}
\left\langle \bar{\psi}\psi \right\rangle _{a} &=&\langle \bar{\psi}\psi
\rangle _{0,\mathrm{ren}}+\frac{q}{2\pi ^{2}}\sum_{n=0}^{\infty }\sum_{p=\pm
1}\int_{m}^{\infty }dx\frac{x}{\sqrt{x^{2}-m^{2}}}\mathrm{Re}\left\{ \frac{%
I_{n_{p}}^{\left( a\right) }\left( ax\right) }{K_{n_{p}}^{\left( a\right)
}\left( ax\right) }\right.  \notag \\
&&\times \left. \left[ \left( sm+i\sqrt{x^{2}-m^{2}}\right)
K_{n_{p}}^{2}\left( rx\right) -\left( sm-i\sqrt{x^{2}-m^{2}}\right)
K_{n_{p}+1}^{2}\left( rx\right) \right] \right\} ,  \label{FCa}
\end{eqnarray}%
where $n_{p}$ is defined by (\ref{np}). It can be seen that in the special
case $s=1$, $\lambda _{a}=1$ this expression coincides with the result from
\cite{Bell11}. The condensate given by (\ref{FCa}) changes the sign under
the replacement $(s,\lambda _{a})\rightarrow (-s,-\lambda _{a})$. Combining
this property with the corresponding behaviour of the last term in (\ref{FC2}%
), we conclude that the FC $\left\langle \bar{\psi}\psi \right\rangle $ in
the region $a\leqslant r\leqslant b$ changes the sign under the
transformation%
\begin{equation}
(s,\lambda _{a},\lambda _{b})\rightarrow (-s,-\lambda _{a},-\lambda _{b}).
\label{Trans}
\end{equation}

For a massless field the FC in the boundary-free geometry vanishes and the
single edge induced contribution is simplified to (see also \cite{Bell11}
for the boundary condition with $\lambda _{a}=1$)
\begin{equation}
\left\langle \bar{\psi}\psi \right\rangle _{a}=-\frac{\lambda _{a}q}{2\pi
^{2}a^{2}}\sum_{n=0}^{\infty }\sum_{p=\pm 1}\int_{0}^{\infty }dx\,\frac{%
K_{n_{p}}^{2}\left( xr/a\right) +K_{n_{p}+1}^{2}\left( xr/a\right) }{%
K_{n_{p}+1}^{2}\left( x\right) +K_{n_{p}}^{2}\left( x\right) }.
\label{FCam0}
\end{equation}%
In this special case the total FC on a conical ring takes the form%
\begin{eqnarray}
\left\langle \bar{\psi}\psi \right\rangle &=&\left\langle \bar{\psi}\psi
\right\rangle _{a}-\frac{q}{2\pi ^{2}}\sum_{n=0}^{\infty }\sum_{p=\pm
1}\int_{0}^{\infty }dx\,x\,\mathrm{Im}\left[ \frac{K_{n_{p}}^{(b)}\left(
bx\right) }{K_{n_{p}}^{(a)}\left( ax\right) }\right.  \notag \\
&&\times \left. \frac{G_{n_{p},n_{p}}^{\left( a\right) 2}\left( ax,rx\right)
+G_{n_{p},n_{p}+1}^{\left( a\right) 2}\left( ax,rx\right) }{%
G_{n_{p}}^{(ab)}(ax,bx)}\right] ,  \label{FC2m0}
\end{eqnarray}%
where now%
\begin{equation}
f_{n_{p}}^{(u)}(z)=\delta _{f}zf_{n_{p}+1}\left( z\right) +i\lambda
_{u}n_{u}zf_{n_{p}}\left( z\right) .  \label{fum0}
\end{equation}%
Of course, for a massless field the FC does not depend on the parameter $s$.
The zero FC for a massless field, realizing one of the irreducible
representations of the Clifford algebra and propagating on a conical space
without boundary, is related to the time-reversal ($T$-)symmetry of the
model. The presence of the edges gives rise to nonzero FC and, hence, breaks
the $T$-symmetry. This mechanism of $T$-symmetry breaking for planar
fermionic systems have been discussed in \cite{Berr87}. The symmetry
breaking was interpreted semiclassically in terms of the phases accumulated
by the waves travelling along closed geodesics inside a bounded region and
reflected from the boundary.

In the representation (\ref{FC2}) for the FC on a conical ring with edges $%
r=a$ and $r=b$, the part corresponding to a cut cone with $a\leqslant
r<\infty $ is explicitly separated. An alternative representation, where the
part corresponding to a cone with finite radius $b$ is extracted, is
obtained from (\ref{FC2}) using the identity%
\begin{eqnarray}
&&\frac{I_{n_{p}}^{(a)}(ax)}{K_{n_{p}}^{(a)}(ax)}K_{\nu }^{2}(y)+\frac{%
K_{n_{p}}^{(b)}(bx)}{K_{n_{p}}^{(a)}(ax)}\frac{G_{n_{p},\nu }^{(a)2}(ax,y)}{%
G_{n_{p}}^{(ab)}(ax,bx)}  \notag \\
&&\qquad =\frac{K_{n_{p}}^{(b)}(bx)}{I_{n_{p}}^{(b)}\left( bx\right) }I_{\nu
}^{2}\left( y\right) +\frac{I_{n_{p}}^{(a)}(ax)}{I_{n_{p}}^{(b)}(bx)}\frac{%
G_{n_{p},\nu }^{(b)2}(bx,y)}{G_{n_{p}}^{(ab)}(ax,bx)},  \label{Id3}
\end{eqnarray}%
with $\nu =n_{p},n_{p}+1$. Separating the contributions coming from the
first term in the right-hand side, the FC in the region $a\leqslant
r\leqslant b$ is presented in the form
\begin{eqnarray}
\left\langle \bar{\psi}\psi \right\rangle &=&\left\langle \bar{\psi}\psi
\right\rangle _{b}+\frac{q}{2\pi ^{2}}\sum_{n=0}^{\infty }\sum_{p=\pm
1}\int_{m}^{\infty }dx\frac{x}{\sqrt{x^{2}-m^{2}}}\mathrm{Re}\left\{ \frac{%
I_{n_{p}}^{(a)}(ax)/I_{n_{p}}^{(b)}(bx)}{G_{n_{p}}^{(ab)}(ax,bx)}\right.
\notag \\
&&\times \left. \left[ \left( sm+i\sqrt{x^{2}-m^{2}}\right)
G_{n_{p},n_{p}}^{\left( b\right) 2}\left( bx,rx\right) -\left( sm-i\sqrt{%
x^{2}-m^{2}}\right) G_{n_{p},n_{p}+1}^{\left( b\right) 2}\left( bx,rx\right) %
\right] \right\} ,  \label{FC3}
\end{eqnarray}%
where%
\begin{eqnarray}
\left\langle \bar{\psi}\psi \right\rangle _{b} &=&\langle \bar{\psi}\psi
\rangle _{0,\mathrm{ren}}+\frac{q}{2\pi ^{2}}\sum_{n=0}^{\infty }\sum_{p=\pm
1}\int_{m}^{\infty }dx\frac{x}{\sqrt{x^{2}-m^{2}}}\mathrm{Re}\left\{ \frac{%
K_{n_{p}}^{(b)}(bx)}{I_{n_{p}}^{(b)}\left( bx\right) }\right.  \notag \\
&&\times \left. \left[ (sm+i\sqrt{x^{2}-m^{2}})I_{n_{p}}^{2}\left( rx\right)
-(sm-i\sqrt{x^{2}-m^{2}})I_{n_{p}+1}^{2}\left( rx\right) \right] \right\} .
\label{FCb}
\end{eqnarray}%
In the limit $a\rightarrow 0$ and for $|\alpha _{0}|<1/2$ the second term in
the right-hand side of (\ref{FC3}) tends to zero as $a^{q(1-2|\alpha _{0}|)}$
whereas the first term does not depend on $a$. This allows to interpret the
part $\left\langle \bar{\psi}\psi \right\rangle _{b}$ as the FC on a cone $%
0\leqslant r\leqslant b$ for a field obeying the boundary condition (\ref%
{BCs}) on a single circular boundary at $r=b$. With this interpretation, the
last term in (\ref{FC3}) corresponds to the contribution when we
additionally add the boundary at $r=a$ with the respective boundary
condition from (\ref{BCs}). In the special case $s=1$, $\lambda _{b}=1$ the
FC (\ref{FCb}) coincides with the result derived in \cite{Bell11}.

For a massless field, from (\ref{FC3}) we get the following alternative
representation for the FC:%
\begin{eqnarray}
\left\langle \bar{\psi}\psi \right\rangle &=&\left\langle \bar{\psi}\psi
\right\rangle _{b}-\frac{q}{2\pi ^{2}}\sum_{n=0}^{\infty }\sum_{p=\pm
1}\int_{m}^{\infty }dx\,x\,\mathrm{Im}\left[ \frac{I_{n_{p}}^{(a)}(ax)}{%
I_{n_{p}}^{(b)}(bx)}\right.  \notag \\
&&\times \left. \frac{G_{n_{p},n_{p}}^{\left( b\right) 2}\left( bx,rx\right)
+G_{n_{p},n_{p}+1}^{\left( b\right) 2}\left( bx,rx\right) }{%
G_{n_{p}}^{(ab)}(ax,bx)}\right] ,  \label{FC3m0}
\end{eqnarray}%
with the single edge contribution
\begin{equation}
\left\langle \bar{\psi}\psi \right\rangle _{b}=-\frac{\lambda _{b}q}{2\pi
^{2}b^{2}}\sum_{n=0}^{\infty }\sum_{p=\pm 1}\int_{0}^{\infty }dx\,\frac{%
I_{n_{p}}^{2}\left( xr/b\right) +I_{n_{p}+1}^{2}\left( xr/b\right) }{%
I_{n_{p}}^{2}\left( x\right) +I_{n_{p}+1}^{2}\left( x\right) }.
\label{FCbm0}
\end{equation}%
The latter is negative for the boundary condition with $\lambda _{b}=1$ and
positive for the condition with $\lambda _{b}=-1$.

The FC in (\ref{FC2m0}) diverges on the edges $r=a,b$. The divergence at $%
r=a $ comes from the single boundary part $\left\langle \bar{\psi}\psi
\right\rangle _{a}$ in the representation (\ref{FC2}) and the divergence on
the edge $r=b$ comes from the term $\left\langle \bar{\psi}\psi
\right\rangle _{b}$ in (\ref{FC3}). In order to find the leading term in the
asymptotic expansion over the distance from the edge at $r=u$, $u=a,b$, we
note that for $|r/u-1|\ll 1$ the dominant contribution in the edge-induced
parts $\left\langle \bar{\psi}\psi \right\rangle _{u}$ (the last terms in (%
\ref{FCa}) and (\ref{FCb})) come from large values of $x$ and $n$. By using
the uniform asymptotic expansions for the modified Bessel functions, to the
leading order we get%
\begin{equation}
\left\langle \bar{\psi}\psi \right\rangle \approx -\frac{\lambda _{u}}{8\pi
(r-u)^{2}}.  \label{FCnear}
\end{equation}%
In deriving this result we have additionally assumed that $m|r-u|\ll 1$.
Near the edges the leading term does not depend on the mass, on the magnetic
flux and on the angle deficit of the conical geometry. It is of interest to
note that the vacuum expectation values of the charge and current densities
are finite on the ring edges \cite{Bell20}.

\begin{figure}[tbph]
\begin{center}
\epsfig{figure=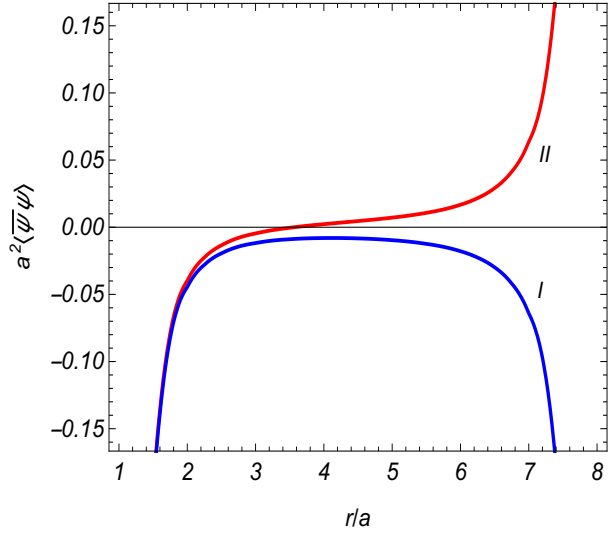,width=8cm,height=7cm}
\end{center}
\caption{The radial dependence of the FC for a massless field on a conical
ring with the parameters $b/a=8$, $q=1.5$, $\protect\alpha _{0}=1/4$. The
graphs I and II correspond to the sets $(\protect\lambda _{a},\protect%
\lambda _{b})=(1,1)$ and $(\protect\lambda _{a},\protect\lambda _{b})=(1,-1)$%
, respectively.}
\label{fig2}
\end{figure}

In Figure \ref{fig2} we display the FC for a massless fermionic field on a
conical ring as a function of the radial coordinate. The graphs are plotted
for $b/a=8$, $q=1.5$ and $\alpha _{0}=1/4$. The curves I and II correspond
to the boundary conditions on the edges with $(\lambda _{a},\lambda
_{b})=(1,1)$ and $(\lambda _{a},\lambda _{b})=(1,-1)$, respectively. The
graphs for the remaining combinations of the set $(\lambda _{a},\lambda
_{b}) $ are obtained by taking into account the property that for a massless
field the FC changes the sign under the replacement $(\lambda _{a},\lambda
_{b})\rightarrow (-\lambda _{a},-\lambda _{b})$. In the case I the FC is
negative everywhere. For the case II the condensate is negative near the
edge $r=a$ and positive near $r=b$. This behavior is in accordance with the
asymptotic estimate (\ref{FCnear}).

\begin{figure}[tbph]
\begin{center}
\epsfig{figure=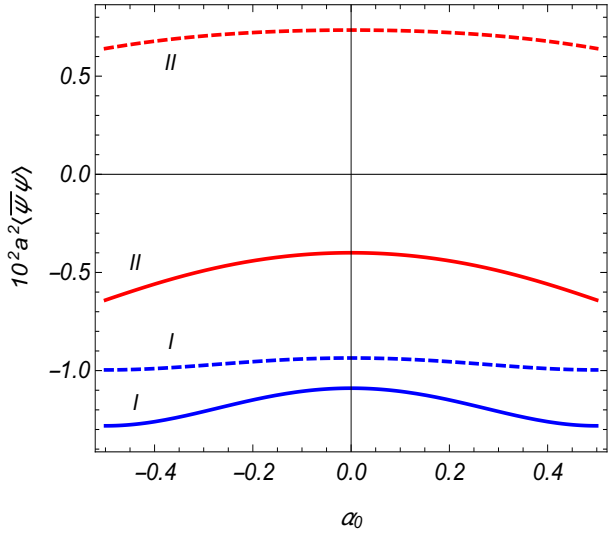,width=8cm,height=7cm}
\end{center}
\caption{The FC versus the parameter $\protect\alpha _{0}$ for a massless
field. The graphs are plotted for $b/a=8$, $q=1.5$, $r/a=3$ (full curves)
and $r/a=5$ (dashed curves). }
\label{fig3}
\end{figure}

The dependence of the FC on the parameter $\alpha _{0}$ is depicted in
Figure \ref{fig3} for a massless field and for the parameters $b/a=8$ and $%
q=1.5$. The full and dashed curves correspond to $r/a=3$ and $r/a=5$. As in
Figure \ref{fig2}, the graphs I and II are for the sets $(\lambda
_{a},\lambda _{b})=(1,1)$ and $(\lambda _{a},\lambda _{b})=(1,-1)$,
respectively. The FC is continuous at half-integer values of the ratio of
the magnetic flux to the flux quantum. The corresponding derivative for the
case $(\lambda _{a},\lambda _{b})=(1,1)$ is continuous as well. For the
boundary conditions with $(\lambda _{a},\lambda _{b})=(1,-1)$ the derivative
of the FC with respect to the magnetic flux is discontinuous for
half-integer values of the ratio of the magnetic flux to the flux quantum.

\begin{figure}[tbph]
\begin{center}
\epsfig{figure=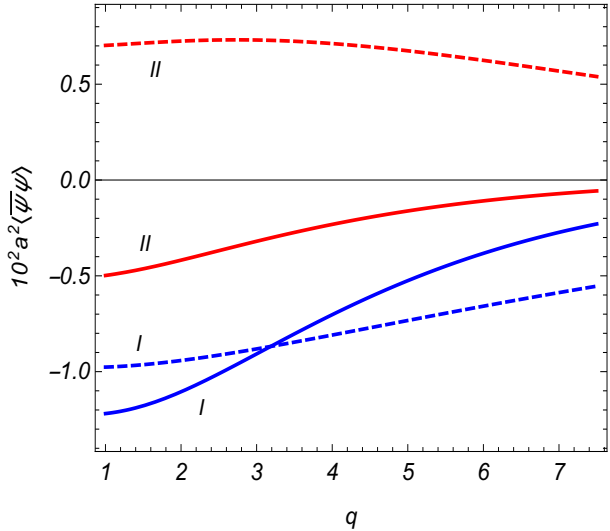,width=8cm,height=7cm}
\end{center}
\caption{The dependence of the FC on the planar angle deficit of the conical
space for $\protect\alpha _{0}=1/4$. The values of the remaining parameters
are the same as those for Figure \protect\ref{fig3}. }
\label{fig4}
\end{figure}

Figure \ref{fig4} displays the FC as a function of the parameter $q$,
determining the planar angle deficit for conical geometry. The graphs are
plotted for a massless field and for the values of the parameters $b/a=8$, $%
\alpha _{0}=1/4$, $r/a=3$ (full curves) and $r/a=5$ (dashed curves). As
before, the curves I and II correspond to $(\lambda _{a},\lambda _{b})=(1,1)$
and $(\lambda _{a},\lambda _{b})=(1,-1)$, respectively.

\begin{figure}[tbph]
\begin{center}
\epsfig{figure=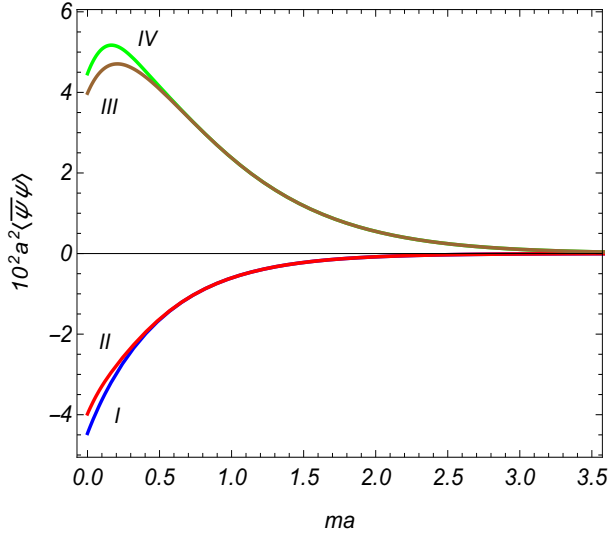,width=8cm,height=7cm}
\end{center}
\caption{The FC as a function of the mass for $s=1$ and for fixed values $%
b/a=8$, $q=1.5$, $r/a=2$, $\protect\alpha _{0}=1/4$. The separate graphs
correspond to different combinations of the boundary conditions on the ring
edges. }
\label{fig5}
\end{figure}

All the graphs above were plotted for a massless field. In order to see the
effects of finite mass, in Figure \ref{fig5} we depicted the dependence of
the FC on the dimensionless parameter $ma$ for $s=1$, $b/a=8$, $q=1.5$, $%
r/a=2$ and $\alpha _{0}=1/4$. The curves I,II,III,IV correspond to the sets
of discrete parameters $(\lambda _{a},\lambda _{b})=(1,1)$, $(1,-1)$, $%
(-1,1) $ and $(-1,-1)$, respectively. The graphs for $s=-1$ are obtained
from those in Figure \ref{fig5} by taking into account that the FC changes
the sign under the transformation (\ref{Trans}). As seen, the dependence on
the mass, in general, is not monotonic. Of course, as we could expect the FC
tends to zero for large values of the mass.

In the discussion above we have considered the simplest configuration of
external gauge field which can be interpreted in terms of the magnetic flux
threading the conical ring. The magnetic field is zero on the ring, where
the fermion field is localized, and its influence is purely topological. New
interesting effects in (2+1)-dimensional fermionic models appear in the
presence of magnetic fields directly interacting with fermions. In
particular, the formation of the FC and chiral symmetry breaking have been
studied extensively in the literature (for a recent review see \cite{Mira15}%
). These investigations have been done within the framework of models with
four-fermion and/or gauge interactions. They have demonstrated that magnetic
fields serve as a catalyst of chiral symmetry breaking and the latter occurs
even in the limit of weak gauge couplings. In some models the FC is related
to the gauge field condensate $\langle F_{\mu \nu }F^{\mu \nu }\rangle $. An
example is the relation $\left\langle \bar{\psi}\psi \right\rangle \sim
-\langle F_{\mu \nu }F^{\mu \nu }\rangle /m_{\psi }$ ($F_{\mu \nu }$ is the
gluon field strength tensor and $m_{\psi }$ is the mass of the $\psi $%
-quark) between the quark and gluon condensates in quantum chromodynamics,
valid in the heavy quark limit \cite{Shif79}. This relation gives the
leading order term in the expansion over $1/m_{\psi }$. The gauge field
condensate can also be formed for abelian gauge fields (see, for example,
\cite{Saha20}\ for the formation of photon condensate in braneworld models
on the AdS bulk).

As it has been mentioned before, in the model under consideration with
boundary conditions (\ref{BCs}) for the eigenvalues of the radial quantum
number one has $\gamma ^{2}+m^{2}>0$ and the vacuum state is stable.
However, the inclusion of fermion interactions may lead to instabilities, as
a result of which phase transitions take place. The FC appears as an order
parameter in those transitions. In particular, the phase transitions, the
chiral symmetry breaking and dynamical mass generation within the framework
of (2+1)-dimensional Nambu--Jona-Lasinio (NJL) (or Gross-Neveu)-type models
with four-fermion interactions have been previously discussed in the
literature (see, for example, \cite%
{Mira15,Klim88,Rose89,Seme89,Eber16,Eber19} and the references \cite%
{Drut08,Juri09,Herb09} for applications in graphene). The influence of
additional boundary conditions on the fermionic field, induced by the
presence of boundaries or by compactification of spatial dimensions, was
investigated as well (see, for instance, \cite%
{Flac12,Flac17,Vita99,Eber08,Abre09,Eber10,Khan12} and references therein).
In particular, it has been shown that those conditions may either reduce or
enlarge the chiral breaking region. In some cases the compactification may
exclude the possibility for the dynamical symmetry breaking.

\section{Fermion condensate in P- and T-symmetric models}

\label{sec:PTsym}

For a fermion field $\psi (x)$ in two spatial dimensions, realizing one of
the irreducible representations of the Clifford algebra, the term $m\bar{\psi%
}\psi $ in the corresponding Lagrangian density is not invariant with
respect to the parity ($P$) and time-reversal ($T$) transformations. The $P$%
- and $T$-symmetries can be restored considering models involving two fields
$\psi _{(+1)}$ and $\psi _{(-1)}$ realizing inequivalent irreducible
representations and having the same mass. The corresponding Lagrangian
density is given by $L=\sum_{s=\pm 1}L_{(s)}$ with the separate terms from (%
\ref{Ls}). We assume that the fields obey the boundary conditions (\ref{BC})
on the ring edges. The total FC is presented in two equivalent forms, $%
\sum_{s=\pm 1}\langle \bar{\psi}_{(s)}\psi _{(s)}\rangle $ and $\sum_{s=\pm
1}\langle \bar{\psi}_{(s)}^{\prime }\psi _{(s)}^{\prime }\rangle $. An
equivalent representation of the model is obtained combining the
two-component fields in a single 4-component spinor $\Psi =(\psi
_{(+1)},\psi _{(-1)})^{T}$ with the Lagrangian density
\begin{equation}
L=\bar{\Psi}(i\gamma _{(4)}^{\mu }D_{\mu }-m)\Psi ,  \label{Lag3}
\end{equation}%
where the $4\times 4$ Dirac matrices are given by $\gamma _{(4)}^{\mu
}=I\otimes \gamma ^{\mu }$ for $\mu =0,1$, and $\gamma _{(4)}^{2}=\sigma
_{3}\otimes \gamma ^{2}$ with $\sigma _{3}$ being the Pauli matrix. For the
corresponding FC one has the standard expression $\langle \bar{\Psi}(x)\Psi
(x)\rangle $. The boundary conditions on the edges $r=a,b$ are rewritten as
\begin{equation}
\left( 1+i\Lambda _{r}n_{\mu }\gamma _{(4)}^{\mu }\right) \Psi (x)=0,
\label{BCPsi}
\end{equation}%
with $\Lambda _{r}=\mathrm{diag}(\lambda _{r}^{(+1)},\lambda _{r}^{(-1)})$.
Alternatively, we can introduce the spinor $\Psi ^{\prime }=(\psi
_{(+1)}^{\prime },\psi _{(-1)}^{\prime })^{T}$ and the set of gamma matrices
$\gamma _{(4)}^{\prime \mu }=\sigma _{3}\otimes \gamma ^{\mu }$. For the
corresponding Lagrangian density one gets $L=\bar{\Psi}^{\prime }(i\gamma
_{(4)}^{\prime \mu }D_{\mu }-m)\Psi ^{\prime }$ and for the FC $\langle \bar{%
\Psi}^{\prime }(x)\Psi ^{\prime }(x)\rangle $. Now the boundary conditions
take the form $\left( 1+i\Lambda _{r}n_{\mu }\gamma _{(4)}^{\prime \mu
}\right) \Psi ^{\prime }(x)=0$. The latter has the same form as (\ref{BCPsi}%
), though with different representation of the gamma matrices.

Note that by adding to the set of the gamma matrices $\gamma _{(4)}^{\mu }$,
$\mu =0,1,2$, the Dirac matrix $\gamma _{(4)}^{3}$ we get the set $\gamma
_{(4)}^{\mu }$, $\mu =0,1,2,3$, that obeys the Clifford algebra in
(3+1)-dimensional spacetime. Now we can construct the chiral $\gamma
_{(4)}^{5}=i\prod_{\mu =0}^{3}\gamma _{(4)}^{\mu }$ matrix which
anticommutes with the $\gamma _{(4)}^{\mu }$, $\mu =0,1,2,3$, matrices and
realizes the chiral transformation $\Psi _{\mathrm{ch}}(x)\rightarrow
e^{i\chi _{\mathrm{ch}}\gamma _{(4)}^{5}}\Psi (x)$ with a phase $\chi _{%
\mathrm{ch}}$. For a massless field the Lagrangian density (\ref{Lag3}) is
invariant under the chiral transformation but the boundary condition (\ref%
{BCPsi}) is not. In the literature bag models for hadrons have been
considered with boundary conditions invariant under the chiral
transformation (chiral bag models, for reviews see \cite%
{Zahe86,Hosa96,Nova96}). In those models the chiral symmetry is restored by
introducing a chiral field that is coupled with the quarks at the bag
surface. The chiral field can be expressed in terms of the isovector pion
field. In chiral bag models the boundary condition on the fermionic field at
the bag surface has the form $\left( e^{iw\gamma _{(4)}^{5}}+in_{\mu }\gamma
_{(4)}^{\mu }\right) \Psi (x)=0$, where the coefficient $w$ in the exponent
is expressed through the pion field on the bag surface.

Let us consider different combinations of the boundary conditions for the
fields $\psi _{(+1)}$ and $\psi _{(-1)}$. First we assume that $\lambda
_{u}^{(+1)}=\lambda _{u}^{(-1)}$, $u=a,b$. For the coefficients in the
boundary conditions for the fields $\psi _{(+1)}^{\prime }$ and $\psi
_{(-1)}^{\prime }$ one gets $\lambda _{u}^{(-1)\prime }=-\lambda
_{u}^{(+1)\prime }$. From here we conclude that the condensates $\langle
\bar{\psi}_{(+1)}\psi _{(+1)}\rangle $ and $\langle \bar{\psi}_{(-1)}\psi
_{(-1)}\rangle $ are obtained from the formulas in the previous sections
taking $s=1$, $\lambda _{u}=\lambda _{u}^{(+1)}$ and $s=-1$, $\lambda
_{u}=-\lambda _{u}^{(+1)}$, respectively. If the parameter $\chi $ in the
condition (\ref{qpc}) and the charges $e$ are the same for the fields $\psi
_{(+1)}$ and $\psi _{(-1)}$, then the parameter $\alpha $ is the same as
well. Now, recalling that the FC discussed in the previous section, changes
the sign under the replacement $(s,\lambda _{u})\rightarrow (-s,-\lambda
_{u})$, we see that the total fermionic condensate vanishes. This means that
in the model at hand with two fields and with the parameters in the boundary
conditions $\lambda _{u}^{(+1)}=\lambda _{u}^{(-1)}$ the Casimir
contributions induced by the edges do not break the parity and time-reversal
symmetries. In the second case with $\lambda _{u}^{(+1)}=-\lambda
_{u}^{(-1)} $, the fields $\psi _{(+1)}$ and $\psi _{(-1)}$ obey different
boundary conditions, whereas for the fields $\psi _{(+1)}^{\prime }$ and $%
\psi _{(-1)}^{\prime }$ the boundary conditions are the same. In this case
the total FC is nonzero and the parity and time-reversal symmetries are
broken by the boundary conditions. Note that, the nonzero FC may appear in
the first case as well if the masses or the phases $\chi $ for separate
fields are different. Hence, the edge-induced effects provide a mechanism
for time-reversal symmetry breaking in the absence of magnetic fields.

Among the interesting condensed matter realizations of fermionic models in
(2+1)-dimensional spacetime is graphene. For a given spin degree of freedom,
the effective description of the long-wavelength properties of the
electronic subsystem is formulated in terms of 4-component fermionic field
\begin{equation}
\Psi =(\psi _{+,A},\psi _{+,B},\psi _{-,A},\psi _{-,B})^{T}.
\label{Psigraph}
\end{equation}%
Two 2-component spinors $\psi _{+}=(\psi _{+,AS},\psi _{+,BS})$ and $\psi
_{-}=(\psi _{-,AS},\psi _{-,BS})$ correspond to two inequivalent points $%
\mathbf{K}_{+}$ and $\mathbf{K}_{-}$ at the corners of the hexagonal
Brillouin zone for the graphene lattice. The components $\psi _{\pm ,A}$ and
$\psi _{\pm ,B}$ present the amplitude of the electron wave function on the
triangular sublattices $A$ and $B$. The Lagrangian density for the field $%
\Psi $ is given as (in standard units)%
\begin{equation}
L_{\mathrm{g}}=\bar{\Psi}[i\hbar \gamma _{(4)}^{0}\partial _{t}+i\hbar
v_{F}\sum_{l=1,2}\gamma _{(4)}^{l}(\nabla _{l}+ieA_{l}/\hbar c)-\Delta ]\Psi
,  \label{Lgraphene}
\end{equation}%
where $c$ is the speed of light, $v_{F}\approx 7.9\times 10^{7}$ cm/s is the
Fermi velocity, and $\Delta $ is the energy gap in the spectrum. The spatial
components of the covariant derivative are expressed as $D_{l}=\nabla
_{l}+ieA_{l}/\hbar c$ with $e$ being the electron charge. Various mechanisms
for the generation of the gap, with the range $1\,\mathrm{meV}\lesssim
\Delta \lesssim 1\,\mathrm{eV}$, have been considered in the literature. For
the corresponding Dirac mass and the related Compton wavelength one has $%
m=\Delta /v_{F}^{2}$ and $a_{C}=\hbar v_{F}/\Delta $. The characteristic
energy scale in graphene made structures is given by $\hbar
v_{F}/a_{0}\approx 2.51\,\mathrm{eV}$, where $a_{0}$ is the inter-atomic
distance for the graphene lattice. The fields $\psi _{+}$ and $\psi _{-}$
correspond to the fields $\psi _{(+1)}$ and $\psi _{(-1)}$ in our
consideration above and the Lagrangian density (\ref{Lgraphene}) is the
analog of (\ref{Lag3}). Hence, the parameter $s$ corresponds to the
valley-indices $+$ and $-$ in graphene physics.

For graphitic cones the allowed values of the opening angle are given by $%
\phi _{0}=2\pi (1-n_{c}/6)$, where $n_{c}=1,2,\ldots ,5$. The transformation
properties of the spinor fields under the rotation by the angle $\phi _{0}$
about the cone axis are studied in \cite{Lamm00,Lamm04,Site07,Chak11}. For
odd values of $n_{c}$ the condition that relates the spinors with the
arguments $\phi +\phi _{0}$ and $\phi $ mixes the valley indices by the
matrix $e^{-i\pi n_{c}\tau _{2}/2}$ with the Pauli matrix $\tau _{2}$ acting
on those indices. One can diagonalize the corresponding quasiperiodicity
condition by a unitary transformation. For graphitic cones with even $n_{c}$
the components with different values of the valley-index are not mixed by
the quasiperiodicity condition. The latter corresponds to (\ref{qpc}) with
the inequivalent values $\chi =\pm 1/3$ for the parameter $\chi $. In
accordance with the consideration given above, if the boundary conditions
and the masses for the fields corresponding to different valleys are the
same, the contributions to the FC coming from those fields cancel each other
and the total FC vanishes. However, some mechanisms for the gap generation
in the spectrum break the valley symmetry (an example is the chemical
doping) and the corresponding Dirac masses for the fields $\psi _{+}$ and $%
\psi _{-}$ differ. In this case one has no cancellation and a nonzero total
FC is formed. As it has been mentioned above, the nonzero FC is also
generated by imposing different boundary conditions on the edges of the ring
for the fields corresponding to different valleys. In these cases the
expression of the FC for a given spin degree of freedom is obtained by
combining the formulas given above for separate contributions coming from
different valleys. In the corresponding expressions it is convenient to
introduce the Compton wavelengths $a_{C+}$ and $a_{C-}$ instead of the Dirac
masses $m_{+}$ and $m_{-}$ through the replacements $m_{\pm }u\rightarrow
u/a_{C\pm }$ for $u=a,b,r$.

The boundary conditions for fermions in the effective description of
graphene structures with edges (graphene nanoribbons) depend on the atomic
terminations. For special cases of zigzag and armchair edges those
conditions have been discussed in \cite{Brey06} (for a generalization see
\cite{Arau19}). The equivalence between the boundary conditions considered
in \cite{Berr87} and \cite{Brey06} has been discussed in \cite{Arau19}. The
boundary conditions for more general types of the atomic terminations in
graphene sheets were studied in \cite{Mcca04,Akhm08,Arau19}. The general
boundary conditions contain four parameters.

\section{Conclusion}

\label{sec:Conc}

The FC is an important characteristic of fermionic fields that plays an
important role in discussions of chiral symmetry breaking and dynamical
generation of mass. It appears as an order parameter for the
confinement-deconfinement phase transitions. In the present paper we have
investigated the FC for a (2+1)-dimensional fermionic field localized on a
conical ring with a general value of the planar angle deficit. The
consideration is presented for both inequivalent irreducible representations
of the Clifford algebra. The boundary conditions on the edges of the ring
are taken in the form (\ref{BCs}) with discrete parameters $\lambda _{a}$
and $\lambda _{b}$. As a special case they include the boundary condition
used in MIT bag model of hadrons for confinement of quarks. The mode-sum for
the FC contains summation over the eigenvalues of the radial quantum number $%
\gamma $. The latter are determined from the boundary conditions on the ring
edges and are roots of the transcendental equation (\ref{Eqgam}). Depending
on the values of the discrete parameters $(s,\lambda _{a},\lambda _{b})$,
one can have modes with purely imaginary values of $\gamma $. For those
modes, corresponding to bound states, we have $\gamma ^{2}+m^{2}\geqslant 0$%
. This shows that for boundary conditions under consideration the fermionic
vacuum state is always stable.

For an equivalent representation of the FC, we have applied the generalized
Abel-Plana-type formula (\ref{Sum}) to the series over the eigenvalues of $%
\gamma $. That allowed to extract explicitly the part in the FC\
corresponding to the region $a\leqslant r<\infty $ of a conical space with a
single adge and to present the part induced by the second edge in the form
of the integral that is well adapted for numerical evaluations (last term in
(\ref{FC2})). The first contribution, corresponding to the conical region $%
a\leqslant r<\infty $ (the second edge at $r=b$ is absent), is further
decomposed in the form of the sum of the boundary-free and edge-induced
terms (formula (\ref{FCa})). An alternative representation of the FC on a
conical ring, given by (\ref{FC3}), is obtained by using the identity (\ref%
{Id3}) for the modified Bessel functions. In that representation the part in
the FC is extracted which corresponds to a finite radius cone (with the
radius $b$) and the last term in (\ref{FC3}) is induced by the second edge
at $r=a$, added to that geometry. For a massless field the boundary-free
contribution in the FC vanishes and the nonzero FC is entirely due to the
presence of boundaries (due to the Casimir effect). In this case the
expressions for the edge-induced contributions to the FC are simplified to (%
\ref{FC2m0}) and (\ref{FC3m0}) with the single-edge geometry parts (\ref%
{FCam0}) and (\ref{FCbm0}). The latter are positive for the boundary
condition with $\lambda _{u}<0$ and negative for $\lambda _{u}>0$.

All the separate contributions to the FC on the conical ring are even
periodic functions of the magnetic flux, enclosed by the ring, with the
period equal to the flux quantum. At small distances from the edge at $r=u$,
$u=a,b$, the leading term in the asymptotic expansion over the distance is
given by the simple expression (\ref{FCnear}). The leading term does not
depend on the mass and on the magnetic flux and is positive (negative) for
the boundary condition with $\lambda _{u}<0$ ($\lambda _{u}>0$). For a
massless field the FC in the boundary-free conical geometry vanishes and the
nonzero contributions are purely edge-induced effects. This provides a
mechanism for $T$-symmetry breaking in the absence of magnetic fields.

For a fermionic field realizing one of the irreducible representations of
the Clifford algebra, the mass term in the Lagrangian density is not
invariant under the parity and time-reversal transformations. Invariant
fermionic models are constructed combining two fields corresponding to two
inequivalent irreducible representations. In those models, the total FC is
obtained by summing the contributions coming from the separate fields. The
latter are obtained based on the results presented in section \ref{sec:FC}.
If the parameters $(\chi ,\lambda _{u})$ and the masses for the separate
fields are the same, then the corresponding contributions cancel each other
and the total FC is zero. In this case the Casimir-type contributions do not
break the parity and time-reversal symmetries of the model. If at least one
of the parameters $(\chi ,\lambda _{u},m)$ is different for the fields in
the combined Lagrangian, the total FC is nonzero and the symmetries are
broken. The results obtained in the paper can be applied for the
investigation of the FC in graphitic cones with circular edges. The opening
angle of the latter can be used as an additional parameter to control the
electronic properties. In the long-wavelength approximation these properties
are well described by the Dirac model with appropriate periodicity
conditions with respect to the rotations around the cone axis.

\section*{Acknowledgments}

A.A.S. was supported by the grant No. 20RF-059 of the Committee of Science
of the Ministry of Education, Science, Culture and Sport RA, and by the
"Faculty Research Funding Program" (PMI Science and Enterprise Incubator
Foundation). T.A.P. was supported by the grant No. 20AA-1C005 of the
Committee of Science of the Ministry of Education, Science, Culture and
Sport RA, and by the "Faculty Research Funding Program" (PMI Science and
Enterprise Incubator Foundation).

\end{document}